\documentstyle[aps,twocolumn,epsfig]{revtex}
\begin{document}
\baselineskip=12pt
\wideabs{
\title
 {The transition temperature of the dilute interacting Bose gas}
\author{Gordon Baym,$^{a,b,c}$ Jean-Paul Blaizot,$^b$
Markus Holzmann,$^c$ Franck Lalo\"e,$^c$ and Dominique Vautherin$^d$
}
\address
      {$^a$University of Illinois at Urbana-Champaign,
      1110 W. Green St., Urbana, IL 61801 \\
$^b$Service de Physique Th\'eorique, CE-Saclay, Orme des Merisiers,
   91191 Gif-sur-Yvette, France\\
$^c$LKB and LPS, Ecole Normale Sup\'erieure, 24 r. Lhomond, 75005 Paris,
   France\\
$^d$LPNHE, Case 200, Universit\'es Paris 6/7, 4 Place Jussieu, 75005 Paris,
France}
\date{\today}

\maketitle

\begin{abstract}
    We show that the critical temperature of a uniform dilute Bose gas
increases linearly with the s-wave scattering length describing the repulsion
between the particles.  Because of infrared divergences, the magnitude of the
shift cannot be obtained from perturbation theory, even in the weak coupling
regime; rather, it is proportional to the size of the critical region in
momentum space.  By means of a self-consistent calculation of the
quasiparticle spectrum at low momenta at the transition, we find an estimate
of the effect in reasonable agreement with numerical simulations.
\end{abstract}
}

    Determination of the effect of repulsive interactions on the transition
temperature of a homogeneous dilute Bose gas at fixed density has had a long
and controversial history [1-5].  While \cite{LY} predicted that the first
change in the transition temperature, $T_c$, is of order the scattering length
$a$ for the interaction between the particles, neither the sign of the effect
nor its dependence on $a$ has been obvious.  Recent renormalization group
studies \cite{stoof} predict an increase of the critical temperature.
Numerical calculations by Gr\"uter, Ceperley, and Lalo\"e \cite{GCL97}, and
more recently by Holzmann and Krauth \cite{HM99}, of the effect of
interactions on the Bose-Einstein condensation transition in a uniform gas of
hard sphere bosons, and approximate analytic calculations by Holzmann,
Gr\"uter, and Lalo\"e of the dilute limit \cite{laloe}, have shown that the
transition temperature, $T_c$, initially {\it rises linearly} with $a$.  The
effect arises physically from the change in the energy of low momentum
particles near $T_c$ \cite{laloe}.  Here we analyze the leading order behavior
of diagrammatic perturbation theory, and argue that $T_c$ increases linearly
with $a$.  We then construct an approximate self-consistent solution of the
single particle spectrum at $T_c$ which demonstrates the change in the low
momentum spectrum, and which enables us to calculate the change in $T_c$
quantitatively.

    We consider a uniform system of identical bosons of mass $m$, at
temperature $T$ close to $T_c$ and use finite temperature quantum many-body
perturbation theory.  We assume that the range of the two-body potential is
small compared to the interparticle distance $n^{-1/3}$, so that the potential
can be taken to act locally and be characterized entirely by the s-wave
scattering length $a$.  Thus we work in the limit $a\ll \lambda$, where
$\lambda=\left({2\pi \hbar^2}/{mk_BT}\right)^{1/2}$ is the thermal wavelength.
(We generally use units $\hbar = k_B = 1$.)

    To compute the effects of the interactions on $T_c$, we write the density
$n$ as a sum over Matsubara frequencies $\omega_\nu=2\pi i \nu T$ ($\nu=0,\pm
1,\pm 2,\dots$) of the single particle Green's function, G(k,z):
\begin{eqnarray}
n = -T\sum_\nu \int\frac{d^3k}{(2\pi)^3} G(k,\omega_\nu).
\label{matsu0}
\end{eqnarray}
where
\begin{eqnarray}
G^{-1}(k,z) = z + \mu - \frac{k^2}{2m} - \Sigma(k,z),
\label{gf}
\end{eqnarray}
with $\mu$ the chemical potential.  The Bose-Einstein condensation
transition is determined by the point where $G^{-1}(0,0)=0$, i.e., where
$\Sigma(0,0) = \mu$.

    The first effect of interactions on $\Sigma$ is a mean field term
$\Sigma_{mf} = 2gn,$ where $g=4\pi \hbar^2 a/m$; the factor of two comes from
including the exchange term.  Such a contribution, independent of $k$ and $z$
has no effect on the transition temperature, as it can be simply absorbed in a
redefinition of the chemical potential.  To avoid carrying along such trivial
contributions we define:
\begin{eqnarray}
\frac{\hbar^2}{2m \zeta^{2}}=-(\mu-2gn).
\end{eqnarray}
The quantity $\zeta$ may be interpreted as the mean field correlation
length.  In the mean field approximation, $\zeta$ becomes infinite at $T_c$;
however, in general, it remains finite, and functions here as an infrared
cutoff.

    Because the effects of interactions are weak, one could imagine
calculating the change in $T_c$ in perturbation theory.  However, such
calculations are plagued by infrared divergences.  Power counting arguments
reveal that the leading contribution to the self-energy, $\Sigma(k\ll
\zeta^{-1},0)$, of a diagram of order $a^n$ has the form:
\begin{eqnarray}\label{sigman}
\Sigma_{n}\sim T\left( \frac{a}{\lambda}\right)^2 \left(
\frac{a\zeta}{\lambda^2}
\right)^{n-2}.
\end{eqnarray}
In perturbation theory about the mean field, with the mean field criterion
for the phase transition, $\zeta\to\infty $ at $T_c$, all $\Sigma_{n}$
diverge, starting with a logarithmic divergence at $n=2$.  More generally, the
approach of $\zeta$ towards $\lambda^2/a$ in magnitude signals, according to
the Ginsburg criterion, the onset of the critical region.  Beyond,
perturbation theory breaks down, since all $\Sigma_{n}$ in Eq.~(\ref{sigman})
are of the same order of magnitude.

    Even though the theory is infrared divergent, we can isolate the leading
correction to the change in $T_c$, which, as we show, is of order $a$.  Since
the infrared divergences occur only in terms with zero Matsubara frequencies
we separate, in Eq.~(\ref{matsu0}), the contribution of the $\nu = 0$ terms,
writing
\begin{eqnarray}
n(a,T)
  = -T\int\frac{d^3k}{(2\pi)^3} \left(G_{\nu=0}(k)+ G_{\nu\ne 0}(k)\right),
\label{matsu}
\end{eqnarray}
where $G_{\nu\ne 0}(k)$ is the sum of terms with $\nu \ne 0$.  Similarly
the density of a non-interacting system with condensation temperature T is
given by
\begin{eqnarray}
  n^0(T)  &=& -T\int\frac{d^3k}{(2\pi)^3}
  \left(G_{\nu=0}^0(k)+ G_{\nu\ne 0}^0(k)\right) \nonumber \\
  &=&
  \int \frac{d^3k}{(2\pi)^3}\frac{1}{e^{k^2/2mT}-1} =
     \frac{\zeta(3/2)}{\lambda^3},
\label{matsufree}
\end{eqnarray}
where $\zeta(3/2)=2.612$.  Since the non-zero Matsubara frequencies
regularize the infrared behavior of the momentum integrals, the dependence of
the $\nu\ne 0$ terms in Eq.~(\ref{matsu}) on $a$ is non-singular at $T_c$.
These terms, of order $a^2$ at least, can be neglected.  Thus to order $a$,
\begin{eqnarray}
  n(a,T_c)- n^0(T_c) = -T_c\int\frac{d^3k}{(2\pi)^3} (G_0(k) - G_0^0(k)).
\label{densdiff}
\end{eqnarray}
To calculate the change in $T_c$ at fixed density we equate $n(a,T_c)$ at
$T_c$ with $n^0(T_c^0)$ at the free particle transition temperature $T_c^0$
and observe that $n^0(T_c) = (T_c/T_c^0)^{3/2} n^0(T_c^0)$; thus in lowest
order the change in transition temperature $\Delta T_c = T_c - T_c^0$ is given
by
\begin{eqnarray}
   \frac32 \frac{\Delta T_c}{T_c}n^0(T_c^0) =
   T_c\int\frac{d^3k}{(2\pi)^3}  (G_{\nu=0}(k) - G_{\nu=0}^0(k)).
\label{transdiff}
\end{eqnarray}
where $\Delta T_c = T_c - T_c^0$.  Thus
\begin{eqnarray}
  \frac{\Delta T_c}{T_c} = \frac{4\lambda}{3\pi\zeta(3/2)}\int_0^\infty dk
  \frac{U(k)}{k^2 + U(k)},
\label{delta}
\end{eqnarray}
where $U(k)\equiv 2m(\Sigma_{\nu=0}(k) - \mu)$.

    Equation (\ref{delta}) for the leading correction to the critical
temperature is crucial.  The criterion for spatially uniform condensation is
that $U(0)=0$; above the transition, $U(0)>0$.  At the transition, $k^2+U(k) >
0$ for $k>0$.  At large wavenumbers, $U\to 1/\zeta^2 > 0$, and in the critical
region, as we discuss below, $U$ is also positive.  Although we have not
proved it rigorously, numerical simulations indicate that $U$ is generally
positive for $k>0$, which implies that the integral in Eq.~(\ref{delta}) and
hence $\Delta T_c$ is positive.

    In the critical region, $k<k_c$, where $k_c$ defines the scale of the
critical region in momentum space, $G_{\nu=0}$ has the scaling form
\cite{josephson} $G_{\nu=0}^{-1}(k) = -k^{2-\eta}k_c^\eta F(k\xi)$; $\xi$ is
the coherence length which diverges at $T_c$ as $|T-T_c|^{-\nu}$, and $F$ is a
dimensionless function, with $F(\infty)\sim 1$.  The critical index, $\eta$,
is given to leading order in the $\epsilon = 4-d$ expansion by $\epsilon^2/54$
\cite{zinn}.  At $T_c$, $G_{\nu=0}^{-1}(k) \sim - k^{2-\eta}k_c^\eta$, and
thus $U\sim +k^{2-\eta}$.  Both terms in Eq.~(\ref{densdiff}) give a
contribution of order $k_c$, so that $\Delta T_c/T_c \sim k_c$.  As we shall
see, $k_c\sim a/\lambda^2$, and hence $\Delta T_c/T_c \sim a/\lambda$.

    To study the leading behavior in $a$ quantitatively, we need concentrate
only on the $\nu = 0$ sector where the full finite temperature theory reduces
to a classical field theory \cite{zinn} defined by the action:
\begin{eqnarray}\lefteqn{
    S\{\phi(r)\} = \frac{1}{2mT}\int d^3r
    \left( \frac{}{}\nabla \phi^*(r)\cdot \nabla \phi(r) \right.}\nonumber\\ &
&\left.  +
    \frac{1}{\zeta^2}|\phi(r)|^2
   + 4\pi a (|\phi(r)|^2 - \langle|\phi(r)|^2\rangle)^2 \right) ;
\label{class}
\end{eqnarray}
the probability of a given field configuration entering the computation of
expectation values, is proportional to $e^{-S\{\phi(r)\}}$.

    The classical theory is ultraviolet divergent, but superrenormalizable.
The divergences appear only in the two-loop self-energy,
$\Sigma^{(2)}_{\nu=0}$ -- effectively the second order self-energy written in
terms of the full $G_{\nu=0}$ rather than the zeroth order Green's functions
-- and can be removed by simple renormalization of the mean field coherence
length, $\zeta$.  Since, henceforth, we calculate only in the classical
theory, we drop the subscript $\nu=0$ . The second order self-energy is
\begin{eqnarray}
\Sigma(k) = - 2g^2 \int \frac{d^3 q}{(2\pi)^3} B(q)
              \frac{T}{\epsilon_{{\bf k}- {\bf q}}},
\label{sigsumnew}
\end{eqnarray}
where $\epsilon_k = (k^2 + \zeta^{-2})/2m$, and the ($\nu=0$)
particle-hole bubble,
\begin{eqnarray}
  B(q) =  \int\frac{d^3 p}{(2\pi)^3}
\frac{T}{\epsilon_{p}\epsilon_{{\bf p}+{\bf q}}},
\label{bubb0}
\end{eqnarray}
is given by
\begin{eqnarray}
    B(q) = \frac{2\pi^2\zeta}{T\lambda^4}b(\zeta q);
\end{eqnarray}
$b(x)\to 1/x$ for $x\gg 1$ and $b(0) = 1/\pi$.

    The integral in Eq.~(\ref{sigsumnew}) is logarithmically divergent in the
ultraviolet.  But in the full theory the momentum integrals are cut off by
distribution functions, $f = (e^{k^2/2mT}-1)^{-1}$, and the ultraviolet
behavior is regular.  To control this divergence we introduce an ultraviolet
momentum cutoff, $\Lambda$, in the classical theory, recognizing that it is in
fact effectively determined in the full theory.  Then
\begin{eqnarray}
 2m \Sigma(k) = -32\pi^2\frac{a^2}{\lambda^4}
  \int_0^{\Lambda\zeta}dx x b(x) L(k\zeta,x),
\label{intbL}
\end{eqnarray}
where
\begin{eqnarray}
  L(k\zeta,x) =  \frac{1}{k\zeta}
  \ln\left(\frac{(x+k\zeta)^2+1}{(x-k\zeta)^2+1}\right).
\end{eqnarray}
The divergent part of the integral comes from the large $x$ tail of
$b(x)$, and contributes $-128(a/\lambda^2)^2 \ln (\Lambda\zeta)$ to
$2m\Sigma$.

    More generally we carry out a diagrammatic expansion of $\Sigma$ in terms
of the {\it self-consistent} $\nu=0$ Green's function, defined by $2mG^{-1}(k)
= -k^2 + \zeta^{-2} +2m\Sigma(k,a,G,\Lambda)$. Note that the dependence of
$\Sigma$ on $\zeta$ enters only through the dependence of $\Sigma$ on $G$.  We
define a {\it renormalized} mean field coherence length by
\begin{eqnarray}
 \frac{1}{\zeta_R^2} = \frac{1}{\zeta^2} -
     128\left(\frac{a}{\lambda^2}\right)^2 \ln (\Lambda\zeta_R).
\end{eqnarray}
Then $G^{-1}(k)$ is given by
\begin{eqnarray}
 -2mG^{-1}(k) =  k^2 + \zeta^{-2}_R +2m\Sigma_F(k,a,G),
\end{eqnarray}
where
\begin{eqnarray}
\Sigma_F(k,a,G) =  \Sigma (k,a,G,\Lambda)
     +128\left(\frac{a}{\lambda^2}\right)^2 \ln (\Lambda\zeta_R)
\end{eqnarray}
is independent of $\Lambda$.  As a function of $\zeta_R$, the Green's
function is independent of the cutoff.

    In fact, a simple power counting argument shows that the finite
part of the self-energy has the form
\begin{eqnarray}
\Sigma_F(k,a,G) = \frac{1}{2m\zeta_R^2}\sigma(k\zeta_R,J)
\label{sigmaf}
\end{eqnarray}
where
\begin{eqnarray}
 J = a\zeta_R/\lambda^2.
\end{eqnarray}
To see this structure we note that a term in the self-energy of order
$a^n$ is the product of a dimensionless function of $k\zeta_R$ times the
$\Sigma_n$ of Eq.~(\ref{sigman}), with $\zeta$ replaced by $\zeta_R$
\cite{scale}.

    The criterion for condensation, $\zeta^{-2}_R +2m\Sigma_F(0,a,G) = 0$,
implies that
\begin{eqnarray}
    1 + \sigma(0,J) = 0.
\label{condcrit}
\end{eqnarray}
Since $\sigma (0)$ is a well-behaved function of only the parameter $J$,
Eq.~(\ref{condcrit} determines the critical value of $J=J^*$ for condensation,
a dimensionless number independent of the parameters of the original problem.
At condensation, the renormalized mean field coherence length $\zeta_R$ tends
to infinity as $a \to 0$, with the product $a\zeta_R$ fixed, thus preventing
a perturbative expansion in $a$.

    At condensation $U(k) = (\sigma (k\zeta_R,J^*)+1)/\zeta_R^2$,
and Eq.~(\ref{delta}) implies the change in $T_c$
\begin{eqnarray}
\frac{\Delta T_c}{T_c}= \hspace{180pt}\nonumber\\
 \frac{a}{\lambda}\left(\frac{4}{3\pi\zeta(3/2)}
   \frac{1}{J^*}\int_0^\infty dx
   \frac{\sigma (x,J^*)+1}{x^2 + \sigma (x,J^*)+1}\right).
\label{delta1}
\end{eqnarray}
Since $J^*$ is determined by the condition (\ref{condcrit}), the result
for $\Delta T_c/T_c$ is linear and expected to be positive in $a/\lambda$.

    We turn now to calculating $\Delta T_c$ explicitly within a simple
self-consistent model based on taking only the zero frequency component of the
leading two-loop approximation self-energy, given by Eq.~(\ref{sigsumnew}).
We construct the $\epsilon_p$ as self-consistent quasiparticle energies at the
transition, i.e., solutions of the equation:
\begin{eqnarray}
  G^{-1}(k,\epsilon_k) = 0=\epsilon_k - \frac{k^2}{2m} -
   (\Sigma(k)-\Sigma(0)).
\label{gf1}
\end{eqnarray}

    The low momentum behavior of $\epsilon_k$ is determined by a familiar
argument \cite{PP}.  In order that the integral (\ref{sigsumnew}) converge in
the infrared limit, $\epsilon_k$ must behave, modulo possible logarithmic
corrections, as $\sim k^{\alpha}$, where $\alpha < 2$.  In this case, the term
$k^2/2m$ in (\ref{gf1}) can be neglected at small $k$.  We then expand
$\Sigma(k)$ about $k=0$.  For $1\le\alpha <4/3$ the self-energy is
sufficiently convergent that $\Sigma(k)-\Sigma(0) \sim k^2$ at small $k$, and
thus cannot be the correct self-consistent solution.  For $\alpha$ with $4/3 <
\alpha < 2$ one has $\Sigma(k)- \Sigma(0) \sim +k^{6-3\alpha}$, so we find
self-consistency, $\Sigma(k)- \Sigma(0) \sim k^{\alpha}$, for $\alpha = 3/2$.
We write the small $k$ part of the spectrum as
\begin{eqnarray}
    \epsilon_k = k_c^{1/2} k^{3/2}/2m.
\label{spectr}
\end{eqnarray}
Here $k_c$ is the wavevector around which the $k^{3/2}$ at low $k$ crosses
over to the $k^2/2m$ free-particle behavior.

    To extract the low momentum structure, below the scale $k_c$, we
evaluate the most divergent terms of
\begin{eqnarray}
 \Sigma(k) - \Sigma(0) =
    -2g^2 T\int \frac{d^3 q}{(2\pi)^3} B(q)
         \left(\frac{1}{\epsilon_{\vec k-\vec q}} -
           \frac{1}{\epsilon_{q}}\right);
\label{smallk0}
\end{eqnarray}
at small $k$.  Since the the $k^{3/2}$ structure arises from the small $q$
behavior of the integral; we evaluate the bubble $B(q)$, Eq.~(\ref{bubb0}), at
small $q$ with the spectrum (\ref{spectr}) for $k<k_c$ and $k^2/2m$ for
$k>k_c$.  Then
\begin{eqnarray}
B(q) =   \frac{4m}{\pi\lambda^2 k_c} ( \ln(k_c/q)+c),
\label{B}
\end{eqnarray}
with $c \approx 2+2\ln 2 -\pi/2$ =1.816.  Thus,
\begin{eqnarray}
 \Sigma(k) - \Sigma(0) = \frac{1024\pi}{15m}
 \left(\frac{a}{\lambda^2}\right)^2 \left(\frac{k}{k_c}\right)^{3/2}.
\label{smallk}
\end{eqnarray}
Identifying the right side of Eq.~(\ref{smallk}) with $k_c^{1/2}
k^{3/2}/2m$, we derive
\begin{eqnarray}
  k_c =32\left(\frac{2\pi}{15}\right)^{1/2} \frac{a}{\lambda^2}
   \approx 20.7 \frac{a}{\lambda^2}.
\label{kc}
\end{eqnarray}
As expected, the scale of the unusual low momentum structure is $a/\lambda^2$.

    Let us, for a first quantitative estimate, assume a spectrum at $T_c$ of
the form $\epsilon_k = k_c^{1/2}k^{3/2}/2m$ for $k \ll k_c$, and
$(k^2+k_c^2)/2m$ for $k \gg k_c$.  We smoothly interpolate between these
limits, writing $U(k) = k_c^{1/2}k^{3/2}/\left(1+(k/k_c)^{3/2}\right)$.
Thus $\int dk U/(k^2+U) \simeq 1.2 k_c$, so that with Eq.~(\ref{kc}),
\begin{eqnarray}
\frac{\Delta T_c}{T_c} \simeq 2.9 an^{1/3}.
\end{eqnarray}
By comparison, Gr\"uter, Ceperley, and Lalo\"e \cite{GCL97} find $\Delta
T_c/T_c \approx 0.34 an^{1/3}$, while the more recent calculation of Holzmann
and Krauth yields $\Delta T_c/T_c \approx (2.3\pm 0.25) an^{1/3}$.  The
agreement of the numerical coefficient, given the simplicity of the
approximations in evaluating the effect of interactions on the transition
temperature, is satisfying.  As will be reported in a fuller paper
\cite{club}, this estimate agrees with that derived from the numerical
self-consistent solution of Eq.~(\ref{gf1}).

    The lowest two-loop calculation does not account fully for the
modification of the transition temperature; indeed, at the critical point, all
diagrams become comparable \cite{PPbook,club}.  Consider, for example, summing
the bubbles describing the repeated scattering of the particle-hole pair
in $B$ \cite{largeN}, thus replacing $B$ in Eq.~(\ref{sigsumnew}) by
\begin{eqnarray}
  B_{\rm eff}(q)=\frac{B(q)}{1+2gB(q)},
\end{eqnarray}
where the two accounts for the exchange terms.  The denominator at small
$q$, from Eq.~(\ref{B}), is given by
\begin{eqnarray}
    1+2gB(q) = 1+\frac{32a}{\lambda^2 k_c} (\ln(k_c/q)+c).
\label{denom}
\end{eqnarray}
Since $k_c \sim a/\lambda^2$, the correction is of order unity, and serves
to modify the spectrum, recalculated from Eq.~(\ref{smallk0}) with
(\ref{denom}), from $k^{3/2}$ to $k^{2-\eta}$, with \cite{club} $\eta \simeq
(1/2) - 1/(2c+k_c\lambda^2/16 a) \simeq$ 0.36.

    To estimate $J^*$, we calculate $\Sigma(0)$ from Eq.~(\ref{sigsumnew})
with the 3/2 spectrum and the leading log in $B(q)$, Eq.~(\ref{B}), and
neglect the contribution for $q>k_c$.  Then $\Sigma(0) \simeq -\kappa^2
a^2/2m\lambda^4$, and $\sigma(0,J) \simeq -\kappa^2 J^2$, so that at $T_c$,
$J^* \simeq 1/\kappa =3/(32\sqrt(2+3c))$.  The self-consistent solution of
Eq.~(\ref{gf1}) yields \cite{club} $J^*\simeq 0.07$.

    The modification at $T_c$ of the spectrum of particles at low momenta
should have direct experimental consequences in trapped Bose condensates.
While a $k^2/2m$ particle spectrum yields a flat distribution $v^2 dn/dv$ of
velocities, a more rapidly rising spectrum, e.g., the $k^{3/2}$ discussed
here, depletes the number of low momentum particles.  These effects become
more pronounced with a larger number of particles and flatter traps, as the
level spacing ceases to control the low energy behavior.

    This work was facilitated by the Cooperative Agreement between the
University of Illinois at Urbana-Champaign and the Centre National de la
Recherche Scientifique, and also supported in part by National Science
Foundation Grants No.~PHY94-21309 and PHY98-00978 and the Deutscher
Akademischer Austauschdienst.  LKB and LPS are Unit\'es Associ\'ees au CNRS
UMR 8552 and UMR 8551.


\begin{references}
\bibitem{scale} Were the classical theory not to contain an ultraviolet
divergence, we could have deduced the structure of Eq.~(\ref{sigmaf}) by
making a trivial scaling in the effective action (\ref{class}) of lengths by
$r = \zeta x$, and the field by $\phi(r) = \theta(x) /\lambda \zeta^{1/2}$,
where $x$ and $\theta$ are dimensionless.  This scaling implies that in the
absence of the ultraviolet divergence the classical theory would be a function
only of the dimensionless parameter $a\zeta/\lambda^2$.  After renormalizing
$\zeta$ we find instead that the theory depends only on the parameter $J =
a\zeta_R/\lambda^2$.
\end{references}
\end{document}